%% file: main.tex
\documentclass[sigconf]{acmart}
\usepackage{accsupp}
\newcommand{\rev}[1]{\textcolor{black}{#1}}

\AtBeginDocument{%
  }

\copyrightyear{2026}
\acmYear{2026}
\setcopyright{cc}
\setcctype{by}
\acmConference[DIS '26]{Designing Interactive Systems Conference}{June 13--17, 2026}{Singapore, Singapore}
\acmBooktitle{Designing Interactive Systems Conference (DIS '26), June 13--17, 2026, Singapore, Singapore}
\acmDOI{10.1145/3800645.3813075}
\acmISBN{979-8-4007-2563-0/2026/06}

\begin{document}




\title[Designing for Collective Access]{Designing for Collective Access: In Search of a Solution to Accessible Communication in a Mixed-Ability Non-Profit}






\author{Xinru Tang}
\affiliation{\institution{University of California, Irvine}\city{Irvine}\state{California}\country{USA}}
\email{xinrut1@uci.edu}

\author{Anne Marie Piper}
\affiliation{\institution{University of California, Irvine}\city{Irvine}\state{California}\country{USA}}
\email{ampiper@uci.edu}

\renewcommand{\shortauthors}{Tang and Piper}

\begin{abstract}
As mixed-ability collaboration has become increasingly focal within accessibility research, managing varied, and sometimes conflicting, access needs has become a key consideration in designing for access. When an accessibility feature or practice benefits some people while constraining others, how should designers navigate these trade-offs? This paper responds to this question by analyzing how a mixed-ability nonprofit worked to make communication accessible to its members as it grew from a small blind-focused athletic group to a larger cross-disability organization. Based on a six-month study that combines interviews and field observations, we show that working with conflicting access needs is not just a technical `problem' but a generative process that sparks reflection on technical constraints and preferences, diverse roles and communication norms, and organizational demands. We therefore argue for rethinking ``conflicts'' in access as key sites for ~\rev{revealing power structures and creating opportunities for accountability and repair.}
\end{abstract}



\maketitle
\section{Introduction}
\input{sections/01intro}

\section{Related Work}
\input{sections/02related-work}

\begin{table*}[]
\begin{tabular}{lll}
\hline
\textbf{Identifier} & \textbf{Self-disclosed Disabilities} & \textbf{Roles or Responsibilities}                                  \\ \hline
Athlete-1                  & Amputee                              & Member of the adult program                                        \\ \hline
Athlete-2                  & Blind                                & Member of the adult program                                        \\ \hline
Athlete-3                  & Amputee                              & Former member of the adult program                               \\ \hline
Athlete-4                  & Blind                                & Former member of the adult program                              \\ \hline
Athlete-5                  & Physical disabilities                & Former member of the adult program                               \\ \hline
Parent-1                  & No disabilities                      & Parent of a child member with physical disabilities \\ \hline
Parent-2                  & No disabilities                      & Parent of an autistic child member \\ \hline
Parent-3                  & No disabilities                      & Parent of a child member with Down Syndrome \\ \hline

Staff-1                  & No disabilities                      & Communication coordinator (Intern)             \\ \hline
Staff-2                  & No disabilities                      & \begin{tabular}[l]{@{}l@{}}Outreach manager, responsible for\\scheduling events with external organizations\end{tabular}                             \\ \hline
Staff-3                  & No disabilities                      & Assistant coach of the youth team                        \\ \hline
Board-1                 & No disabilities                      & Board member (therapeutic program director)   \\ \hline
Board-2                  & No disabilities                      & \begin{tabular}[l]{@{}l@{}}Board member (athletic director) \&\\ coach of the adult team\end{tabular}      \\ \hline
Board-3                  & No disabilities                      & Board member (fundraising \& outreach)      \\ \hline
Board-4                  & No disabilities                      & Board member (research \& development)       \\ \hline
Board-5                  &    No disabilities                   & Board member (fundraising \& outreach)       \\ \hline
Board-6                  & Low vision                      & Board member (director)       \\ \hline
\end{tabular}
\caption{Interview participants and their relations to the organization.}
\label{table::participants}
\end{table*}

\BeginAccSupp{method=pdfstringdef, unicode, ActualText={Table 1: This table lists 17 participants in the study, grouped as athletes, parents, staff, and board members. For each participant, three columns are reported: identifier, self-disclosed disabilities, and organizational roles or responsibilities.
Athletes (5 total): Two are amputees, two are blind, and one reports physical disabilities. Their roles include current or former membership in the adult program.
Parents (3 total): None reported disabilities. Each is the parent of a child member with different conditions: physical disabilities, autism, or Down syndrome.
Staff (3 total): None reported disabilities. Their roles include communication coordinator (intern), outreach manager, and assistant coach of the youth team.
Board members (6 total): Five reported no disabilities and one reported low vision. Their responsibilities span therapeutic program direction, athletic direction and coaching, fundraising and outreach, research and development, and general directorship.
Figure 1: A hierarchical chart that maps the organizational structure. At the top level are Board Members. Directly below the board is a single layer labeled Operations. Beneath Operations, the chart divides into two major sections: Executive Staff and Admin (on the left) and Programs (on the right).
Executive Staff and Admin branch: This section branches further into two columns. The first column lists: Secretary (\$), Media Making (intern), Communication Coordinator (Intern), and Accountant. The second column lists: IT, Treasurer, and Equipment Manager (*), and Development.
Programs branch: This section is subdivided into four program areas: Adult, Youth, Recreational (Short Program), and Outreach. Adult program has a Coach (\$). Youth program has a Coach and Assistant Coach. Recreational program has a Coach (*).
Outreach program includes a Manager, Catering, and Hospitality.
Symbols: A dollar sign (\$) indicates that an individual holds multiple roles. An asterisk (*) indicates that the person is also an athlete in the adult program.}}
\begin{figure*}[t]
\centering
\includegraphics[trim=1cm 1.8cm 1cm 3cm, clip, width=\linewidth]{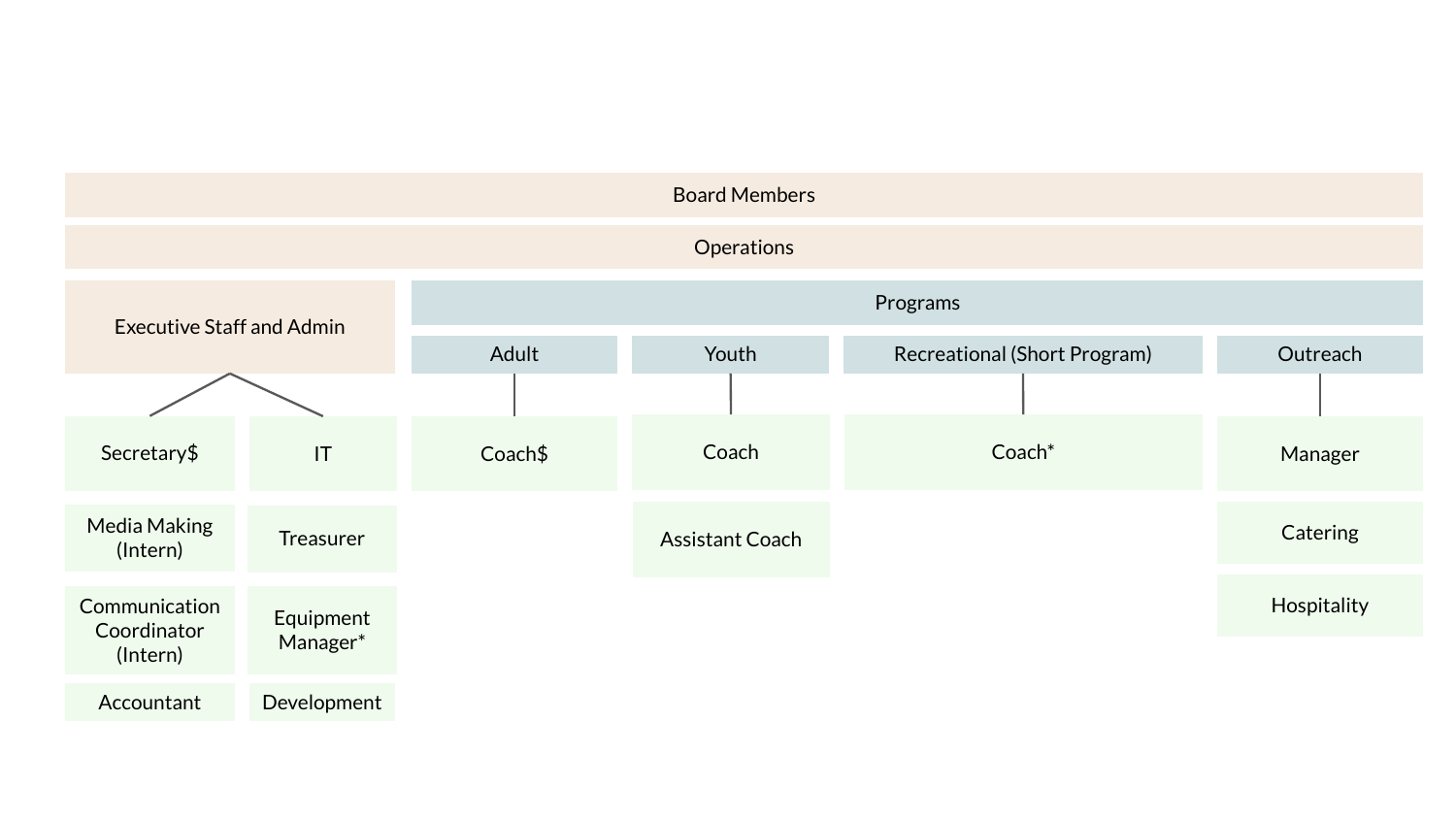}
\caption{Organizational structure of the regular members at our field site during the period of our study. This chart is not comprehensive as it did not include participants of each program, high turnover, and role flexibility. For example, outreach events often rely on volunteers to help with setup, coordination, and on-site assistance. Some individuals hold multiple roles (marked with \$). Asterisks ($\ast$) indicate participants who are also athletes in the adult program.}
\label{fig::org-chart}
\end{figure*}
\EndAccSupp{}

\section{Methods}
\input{sections/03methods}

\section{Findings}
\input{sections/04findings}

\section{Discussion}
\input{sections/05discussion}

\section{Conclusion}
Through our inquiry into the collective access practices at a mixed-ability non-profit, we traced how the organization negotiated the use of communication technologies as it expanded from a small BLV-focused team to a larger, professional, cross-disability organization. Our findings reveal the complexities this organization faced in achieving collective access among its members, which are shaped by its diverse membership, technical constraints, mixed roles and levels of engagement, and evolving organizational demands. Our analysis highlights collective access not as a technical problem with an ideal solution, but as an ongoing process that continually shifts dynamics of inclusion and in which ``conflicts'' serve as key sites of ~\rev{reflection and repair} of relationships and systems that cause power asymmetries.

\begin{acks}
We thank our fieldwork participants for their time, generosity, and knowledge sharing. We are grateful to Kapil Garg for his valuable support throughout this project. We also thank the anonymous reviewers for their constructive feedback on our work. This work was supported by NSF Award \#2326023.
\end{acks}

\bibliographystyle{ACM-Reference-Format}
\bibliography{main}
\end{document}

%% file: sections/01intro.tex
Scholarship in accessible computing, as well as the broader realm of disability activism, often positions improving access or accessibility as the central goal. Subsequently, the longstanding interest among computing researchers and designers has been in designing tools, spaces, and systems that people with disabilities (PWDs)\footnote{We use both people-first and identity-first language to recognize diverse preferences and naming practices within disabled communities.} can use and meet their needs. This pursuit of access has led to extensive research and efforts on developing assistive technologies or accessible artifacts for PWDs~\cite{giraud-2020-making, karim-exploring-2023, haque-topsen-2025, yuksel-human-2020, tang2026disability, yin-malicious-2024}, accessibility guidelines~\cite{w3c-guidelines, zhang2025guidelines}, understanding disabled people's access needs~\cite{goddard-arts-2024, heyko-2021-identifying, tang2023community, tran2026toward, zhang2026access}, and identifying accessibility issues of existing technologies~\cite{lindgren-drones-2024, tang2025everyday, garg2025s, tang2026reimagining}, including collaborative workplace systems~\cite{akter2025beyond, das2019doesn, das2024comes}. Frequently, these efforts result in the development of an improved `solution' to accessibility or design recommendations to make technologies or systems more accessible to a particular group of disabled users. 

However, the turn towards understanding and designing technology for mixed-ability groups (i.e., those comprised of people with varying disabilities and without) introduced additional design complexities~\cite{xiao2024systematic, alharbi2023accessibility, mack2021mixed, yildiz2023virtual, das2019doesn}. In these settings, multiple people's access needs often conflict with each other~\cite{alharbi2023accessibility, mack2021mixed, yildiz2023virtual}, such as varying needs regarding the use of videoconferencing chat and camera features among blind\footnote{~\rev{The definition and operationalization of blindness vary across legal, medical, and functional perspectives~\cite{giudice2018navigating}. We use `blind' for members who use screen readers for digital access, `low vision' for those who primarily use their residual vision, and `BLV' to refer to visual disabilities more broadly, unless for participants who have specified their preferred identities.}} and deaf and hard-of-hearing (DHH) colleagues. Other times, disabled people's access needs may conflict with organizational demands, such as when a blind employee's need to use a screen reader is constrained by restrictive organizational software policies~\cite{apara2025}. As such, the design of assistive technology, or use of certain accessibility features, for one disabled person may in fact be inaccessible and create new problems for others or be at odds with organizational needs~\cite{alharbi2023accessibility, apara2025}. This dilemma, typically framed as ``conflicting access needs''~\cite{mack2021mixed, yildiz2023virtual} or as ``conflicts'' in co-presence~\cite{hofmann2020living}, has turned attention to how collective access, or access in groups, is negotiated and how norms are created.

While understanding collective access as negotiated, and how disabled people themselves do such negotiation work~\cite{yildiz2023virtual, mack2021mixed, das2019doesn}, is a useful step forward, the notion of ``conflicting access needs'' still leaves a dilemma for designers: accommodating one person's needs can come at the expense of others' needs. Even when such accommodations are temporary and collectively negotiated, they can still produce real material and social consequences, often placing disabled people in unequal power relations, for example, requiring them to disclose their needs to their leaders~\cite{das2019doesn, mack2021mixed}. What is more, the framing of ``conflicts'' emphasizes the need to resolve misalignment between multiple disabled bodies or between disabled bodies and non-disabled bodies rather than centering the socio-material realities and politics that create such conflicts in the first place.

To help rethink how to design for conflicting access needs and interrogate the logic that prioritizes designing for resolution of such conflicts, this paper draws on fieldwork with a mixed-ability non-profit organization as they encounter and respond to varied access needs for communication. This organization has spent years `searching' for a solution to make their communication and work accessible to its members. Similar to many studies that seek to identify solutions for accessibility, our research began with the organization articulating to us their desire to build a more accessible communication platform as it expanded from a small blind and low-vision (BLV) athletic group to a larger, professional, cross-disability organization open to people of all ages and (dis)abilities. Because the organization employs and serves both non-disabled members and individuals with a range of disabilities, the selection and use of communication tools for their daily work emerged as a key `problem' since its inception.

Through a six-month study combining field observations and interviews with staff and program participants (n=17), we trace how members of the organization search for an `ideal' communication solution through constant experimentation with new tools and negotiation over members' diverse needs. We find that their ongoing negotiation of access, both in terms of what access means for various members of the organization and how it is achieved, is not simply to find ``better'' technologies or develop solutions to resolve differences in members' needs. Rather, it is a holistic response to wider organizational dynamics including disparate preferences and expectations for technology use; diverse roles and norms for engagement in work; and shifting organizational demands from cross-disability expansion. First, we show that there exists no single `technology' to meet the multiplicity of stakeholder needs and preferences, such as needs for collaborative tools that can support variations in communication structure and diverse accessible workflows. Instead, assemblages of technologies that are (in)accessible to certain people at certain times create an increasingly fractured communication infrastructure. Second, the negotiation of collective access is complicated by heterogeneous roles, expectations, and communication norms across the organization, including varying levels of engagement and mixed working models. ~\rev{These varied expectations for `successful' communication, both technically and in terms of organizational roles,} raised tensions in who should take on the labor to maintain it. Third, we illustrate how collective access is further entangled with changing organizational demands arising from its cross-disability mission, from policy changes to adapting to shifts in disability membership under limited resources. This way, what is often called ``conflicting access needs'' becomes a key driver for reflection and negotiation across a broader network of people, tools, relationships, and hierarchies within and beyond the organization, especially around whose needs might be marginalized and how to repair those exclusions.

Our study makes two primary contributions to the literature on design, accessibility, and HCI. First, we make an empirical contribution through an in-depth analysis of a mixed-ability nonprofit's adoption and use of communication technologies, illustrating the complexities in creating collective access within an organization that serves cross-disability communities and employs both people with and without disabilities. In contrast to prior work on collaboration that centers the experiences of one disabled group (e.g., blind members within blind or sighted teams~\cite{apara2025, das2019doesn, cha2025dilemma}) or role (e.g., employees in mixed-ability teams~\cite{mack2021mixed}), our findings broaden empirical knowledge of mixed-ability collaboration to encompass organizational access across multiple stakeholders, including leadership, operations, and program participants. Second, drawing on critical disability scholarship and disability activism~\cite{hamraie2019crip, leah-collective, dj-sins, titchkosky2011question, dolmage-ableism}, as well as theorizing of access as within HCI and CSCW~\cite{bennett2020care, das2019doesn,  wang2018accessibility, bennett2018interdependence}, we argue that conflicts in collective access are generative rather than merely design `problems' to solve. We discuss what this generative view of conflicts means for designing for collective access.

%% file: sections/02related-work.tex
Our work builds on prior research on the accessibility of collaborative work technologies and dynamics in mixed-ability work more broadly. We also draw on studies of nonprofits' adoption and use of technologies as an essential context for our study.

\subsection{Accessibility of Collaborative Work Technologies}
Research on work accessibility has traditionally focused on accessibility of specific collaborative technologies such as tools and techniques for collaborative writing~\cite{das2019doesn, das2022co11ab, das2022design}, video-conferencing~\cite{alharbi2023accessibility, akter2023if, li2024re-envisioning, mcdonnell2023easier, kim2022deaf, kushalnagar2020teleconference, ang2022online, curtis2025video-calling, wu2024finding}, and digital whiteboards~\cite{das2024comes}. This line of work has revealed widespread accessibility issues in existing technologies that impose additional labor on disabled individuals to navigate them~\cite{branham2015invisible, akter2023if, akter2023if}. For instance, studies have documented various usability issues encountered by screen reader users during virtual meetings, such as challenges in maintaining screen reader focus and navigating complex interfaces~\cite{akter2023if, alharbi2023accessibility}. A recent online survey conducted by Akter et al. with 155 BLV users evaluated the accessibility of 30 commonly used collaboration tools~\cite{akter2025beyond}. The findings reveal that more than half of the respondents reported that accessibility barriers negatively affected their ability to collaborate, perform their jobs, and advance in their careers~\cite{akter2025beyond}. Similarly, research has examined other groups' experiences with work technologies, such as people who are DHH~\cite{mcdonnell2023easier, kim2022deaf, kushalnagar2020teleconference, ang2022online}, who stutter~\cite{li2024re-envisioning, wu2024finding}, or who have chronic illness~\cite{curtis2025video-calling}. Together, these studies have identified a range of accessibility challenges, including inaccurate captioning~\cite{mcdonnell2023easier, li2024re-envisioning, wu2024finding}, insufficient support for sign language~\cite{ang2022online}, and increased fatigue due to extended screen time~\cite{curtis2025video-calling}. In response, researchers have developed systems and techniques to improve these issues~\cite{das2022co11ab, das2022design, mcdonnell2023easier, li2024re-envisioning, lee2022collabally}. Despite these rich insights, much of this work has focused on specific technologies or disabilities, whereas accessibility is far more complex in many real-world situations such as in mixed-ability groups where access is shaped by nuanced interpersonal and organizational dynamics.

\subsection{Accessibility in Mixed-Ability Teams and Organizations}
Recent work has begun to examine dynamics in mixed-ability workplaces more broadly, including within teams that have members with diverse disabilities~\cite{yildiz2023virtual, mack2021mixed}, in interactions with leaders or colleagues~\cite{heyko-2021-identifying, newman2025disclosure}, and across organizations such as in the technology industry~\cite{apara2025}. In contrast to work that emphasizes improving the accessibility of technologies, this line of research often highlights ``norm making'' as central to understanding and creating access in teams, given the challenges of adapting to varied and sometimes even ``conflicting'' access needs among individuals~\cite{yildiz2023virtual, mack2021mixed}. For example, while DHH participants may rely on visual cues, these same cues can distract some neurodivergent participants who need a visually simple environment that minimizes distraction, a situation making it difficult to meet both needs simultaneously~\cite{alharbi2023accessibility, hofmann2020living, das2021towards}. As a result, researchers have advocated for moving beyond individual accessibility to focus on ``collective access'' and resolving ``conflicts'' at the team level~\cite{mack2021mixed, yildiz2023virtual} For example, Mack et al. recommend ``holding a team discussion to establish norms collectively and holistically as a community.''~\cite{mack2021mixed} While crucial, researchers have also noted challenges in building collective access, especially in organizations because of changes in members~\cite{mack2021mixed}, pressures to balance competing values and priorities such as profitability~\cite{apara2025}, maintaining professionalism in front of leaders or colleagues~\cite{mack2021mixed, das2019doesn}, and coordinating with external organizations~\cite{yildiz2023virtual}. All of these factors complicate efforts to practice collective access and are considered key challenges to designing for access in the literature.

\subsection{Nonprofits' Adoption and Use of Technologies}
Our study of mixed-ability work is uniquely situated in the nonprofit context. In the United States, more than 1.5 million nonprofit organizations were registered in fiscal year 2024 \cite{number-ngo}. Many of them operate with disability-centered missions and act as primary service providers and employers for PWD~\cite{ngo-report}, making them key sites for studying how accessibility is understood, implemented, and experienced. Unlike traditional companies with well-established roles and boundaries, nonprofits often feature fluid organizational structures~\cite{tang2015restructuring, voida2011homebrew}, limited resources~\cite{voida2011homebrew}, and insufficient technologies~\cite{voida2011homebrew}. Consequently, systems designed for conventional for-profit businesses are often ill-suited to nonprofits, which rarely have the resources to invest in enterprise applications, hire IT staff, or provide formal training~\cite{tang2015restructuring, voida2011homebrew}. As a result, many nonprofits have to learn to adapt publicly available tools including paperwork to meet their information needs, a practice Voida et al. describe as ``homebrew databases'' \cite{voida2011homebrew}. Beyond operational demands, relationship building and care practices are also vital to non-profits, as many of them depend on volunteers to run their services. Their communication needs therefore extend beyond transactions of money, goods, or services to include expressing gratitude and fostering a sense of community \cite{harmon2017supporting, gui2022community}. Despite the growing literature within HCI on non-profits, a notable gap remains in non-profit organizations that serve disability-related purposes, despite their significant roles in both providing services to and employing PWD~\cite{buettgen2020role, ngo-report}. Our study contributes to this literature with a case of a mixed-ability nonprofit’s adoption and use of communication technologies.

%% file: sections/03methods.tex
Our methods involve six months of field observations (from March to August 2025) coupled with in-depth interviews with 17 community affiliates. Our approach was largely informed by constructivist grounded theory, in which we engaged with iterative and concurrent data collection, coding, and analysis~\cite{charmaz2006constructing}. All study procedures were approved by the IRB of our university and by the organization's leadership.

\subsection{Research Site}
Our field site is a nonprofit organization based in Southern California, USA, which has, for over 15 years, provided competitive and recreational outdoor water activities and services for PWDs. The organization started with a small team of BLV paddlers who were interested in competing in professional races. Over the years, they have expanded to leisure and therapeutic purposes, and started programs open to other disabilities and children. 

At the time of this study, the organization has two main racing programs, one for adults and another for youth and children. Each program has approximately 10 to 20 members. The adult program mainly includes members who are BLV or have physical disabilities, whereas the youth and children's program mainly includes members who are autistic or have other neurodevelopmental disabilities. In addition to its weekly programs, the organization hosts short recreational programs and monthly public outreach events for disabled community members, families, and other disability-focused organizations, serving hundreds of people each year. The staff team has also grown to include six board members and an operating team responsible for coaching, accounting, etc. (refer to Figure \ref{fig::org-chart} for an overview of the organizational structure at the time of this study). ~\rev{While the director identifies as low vision and some roles were taken by BLV athletes, most of the regular staff members and volunteers do not identify disabled}.  

The organization's day-to-day work is shaped by two primary higher-level activities. First, \textbf{routine communication for regular programs}: staff must keep program participants informed about weekly practice schedules and upcoming races. While practice days are fixed each week, coaches need RSVPs before each session to confirm whether practice will proceed and to assign the appropriate equipment based on group size. They also need to share information about upcoming races to gauge interest and tailor training sessions accordingly. Second, \textbf{public events coordination}: the organization manages a variety of public events including corporate team building, open community events, summer camp, and media press. For these events, they need to recruit volunteers, assign roles, and manage logistics such as catering and equipment setup.

Our research started with a conversation with the organizational leadership about communication challenges they have been facing and their perceived need to support their work and streamline the communication tools they use. While the organization has a physical gathering space, i.e., the local aquatic center where they store equipment and launch boats, much of its communication, planning, and coordination work remains distributed online, largely due to the (semi-)voluntary nature of work among a geographically scattered community and the lack of a dedicated place to work like a shared office space. Over the years, they have experimented with various communication platforms to accommodate members' diverse needs, ~\rev{ranging from methods more accessible to BLV members like texting to more structured options like Slack. However, none could meet everyone's needs.} Consequently, they have adopted a range of tools including Slack, WhatsApp, text messaging, and more to maintain their communication and work. While this approach made sure all members involved in communication, it led to fragmented workspaces and communication flows. Our study began with the aim of understanding various stakeholders' communication practices and breakdowns as well as opportunities for enhanced workflows. 

\subsection{Field Observations}
To better understand how the organization communicates and coordinates its activities and to support the interviews, the first author conducted roughly 24 hours of field observations over a period of six months, attending both small group practice sessions and larger community events. Although our primary method was interviews, these observations were essential for understanding the organization’s structure, its activities, the levels of involvement and engagement within this remotely distributed setting, and the community vibe. Additionally, in-person contact remains a crucial part of how members of this organization communicate due to the physical, in-person nature of their services and activities. ~\rev{Field observations also allowed us to engage in informal conversations with individuals for whom time constraints made formal interviews infeasible or remote interviews inaccessible. These conversations in the field allowed us to understand the organization more comprehensively and avoided our understanding skewed toward a small set of members who were available for interviews. In total, we talked to more than 15 additional program participants and staff members, including more athletes who are BLV or have physical disabilities, parents of disabled children, staff members, on-site volunteers, and newcomers to community events.} While in the field, we clarified our research purpose and introduced our team background before continuing conversations with participants who were unfamiliar with us. After each observation session, the lead author wrote detailed field notes for later analysis~\cite{emerson2011writing}. We generated field notes on all observations and conversations with organization members to and include them in our analysis.

\subsection{Interviews}
Participants were initially recruited to participate in in-depth interviews using network sampling methods~\cite{creswell2016qualitative} and then, as our study progressed, we used purposive or theoretical sampling~\cite{charmaz2006constructing}. Leadership staff at the organization assisted in distributing recruitment information and a screener survey with their program participants and within their staff teams. They also invited the lead author to join their Slack workspace, where we sent direct recruitment invitations to staff members (as instructed by the director). 

A total of 17 participants, including program participants and staff members, took part in one-on-one in-depth interviews (see Table~\ref{table::participants} for details), led by the first author. Out of ethical considerations, recruitment was limited to adults able to provide informed consent, excluding adults with neurodevelopmental disabilities who were unable to consent on their own and children under 18. In addition to current members, we also recruited former members to understand the experience of those who have left the organization. The former members left the organization for various reasons, such as joining a club closer to home or exploring other types of water sports. The former members provided valuable context regarding the organization's history and their experiences. ~\rev{To avoid skewing our understanding toward members who were available for interviews, we engaged with members during field observations who did not participate in formal interviews. Including those we met during field observations, we spoke with athletes and youth members with visual, physical, and neurodevelopmental disabilities, as well as members without disabilities.}

Prior to each interview, we shared our research purpose and team background, and emphasized the voluntary nature of the study in the recruitment message, survey screener, and at the beginning of each interview. We conducted the interviews via Zoom, phone calls, or on site, depending on participants' preferences, with informed consent obtained accordingly. During the interviews, we asked participants to describe their involvement at the organization, the tools they have used for communication within the organization, communication breakdowns they experienced, and their thoughts on the future of the organization. Each interview lasted from 30 minutes to 1 hour. We recorded and transcribed the conversations, except for two participants who declined to be recorded. We took detailed notes during these two interviews instead. We offered participants USD\$40 cash or an equivalent amount in a gift card as compensation.

\subsection{Data Analysis}
Our approach is guided by constructivist grounded theory (CGT), in which we engaged in simultaneous data collection and analysis and constant comparison of data to data and data to developing concepts~\cite{charmaz2006constructing}. As part of our process, we conducted ongoing open, inductive coding of the cumulative interview data, wrote analytic memos after each interview, engaged in theoretical sampling, and regularly discussed the resulting themes. We analyzed interview data alongside observational field notes, which supported and enriched our interpretation of the interview data. In addition, we drew on data from the organization’s website and social media to gain a broader understanding of its activities. While the outset of our inquiry was informed by our community partner's need for a technology `solution' to access, as data collection and analysis proceeded, we shifted our analytic frame towards the politics that animate interactions, choices, and norms regarding access rather than any one technology system or artifact. We actively read our data alongside related concepts and theorizations of access, particularly work on collective access practices in mixed-ability work or collaborative engagements (e.g., \cite{mack2021mixed, yildiz2023virtual, das2019doesn, apara2025}) and critical disability scholarship and disability activism that conceptualize access as contested, friction-filled work (e.g., ~\cite{hamraie2019crip, hamraie2019crip, leah-collective, dj-sins, titchkosky2011question}). Through our analytic process, we came to understand that the pursuit of technical `solutions' to access incited critical reflection throughout the organization on desperate needs and preferences among members, diverse roles and communication norms, and shifting organizational demands. Thus, moments of `conflict' in access were generative and created opportunities for the organization to reflect and adapt strategies to ensure that marginalized members remained connected.

\subsection{Positionality and Ethical Considerations}
Our analysis is inherently shaped by our perspectives as computing researchers trained in accessible computing and critical disability scholarship, as well as by our positions within a U.S.-based academic institution. Additionally, the lead author's experiences with disadvantage in traditional competitive sports, as well as her experiences as a non-native English speaker living in an English-speaking country, inform her thinking about what counts as access and inclusion for different stakeholders and from multiple perspectives when entering the field site. While we built relationships with members of the organization's leadership before we started this research and deepened our understanding through ongoing involvement in community events, our interpretations remain shaped by our roles as researchers and observers. Acknowledging access as a political issue, we attended to multiple stakeholders' viewpoints in our analysis, not only the leadership but also the operational staff, program participants, and members who left the organization. Given the small size and close-knit nature of the community we studied, we were careful to protect participants’ privacy. We avoided sharing identifiable stories with other community members and excluded identifying quotes or experiences when reporting our findings.

%% file: sections/04findings.tex
As members of our field site expressed from the outset, understanding and adjusting to people’s diverse access needs for communication has been a key source of challenge in the organization's day-to-day work. Members of the organization depend on a wide range of different devices, including iPhones, Android phones, and flip phones. They also vary in the extent to which they experience certain features of collaboration platforms (e.g., Slack) or modes of communication (e.g., text messaging) as accessible. That is, no tool or practice exists that can fully meet everyone's access needs. What's more, we find that accessible ways of interacting with technology often conflict with the organization's need for coordinating work and disseminating information. Thus, `conflicts' in access give rise to moments of negotiation of what access means and how it is achieved within the organization, spanning a broader network of people, tools, relationships, and hierarchies both within and beyond the organization. Our analysis shows their work of practicing collective access is not simply about finding a `better' tool or suite of tools, nor is it limited to overcoming inaccessible technologies. Instead, `conflicts' in access require a holistic response to disparate preferences and expectations for communication; diverse roles and norms for engagement in work; and shifting organizational demands from cross-disability expansion. 

\subsection{Managing Disparate User Needs and Preferences for Communication}
Much of the discussion of accessible communication within this organization was around the search for tools that can both meet their needs for collaborative work and members’ diverse communication needs and preferences. For years, the team relied primarily on texting, as it was perceived as more accessible to BLV members. However, as the organization grew, texting began showing its limitations as it often led to multiple overlapping group threads, lacked topic-based channels, and made it difficult to organize conversations or track information. Texting also made it harder to share files and documents, and lacks features to store and track information over time. ~\rev{Although the short-term summer program continues to use texting as it is a small and not regular program, the organization has been exploring alternative communication options for its other divisions.} ~\rev{However, the options available were either visually oriented and inaccessible to BLV members, unfamiliar to less tech-savvy users, or hard to manage large volumes of information, with no single tool meeting everyone's needs.} While these explorations did not lead to a singular or `ideal' solution, they prompted important discussions over each tool’s strengths and limitations, how these tools might marginalize some members’ participation, (re)distribute the work of access, and what steps could be taken to address these power asymmetries. All of them constitute part of the ongoing work of access within this organization.

\subsubsection{Exploring Tools' Pros and Cons}
The ongoing discussion has been guided by members' varied preferences for communication. A key conversation took place when the leadership attempted to adopt Slack as the shared workspace, which many believed to have more organized spaces and workflows better suited to the organization's expanding size. However, this plan received mixed response across the organization as blind members had strong preferences for non-visual communication methods. As Board-6 explained the hard situation he faced, 
\begin{quote}
    ``We started using Slack 10 years ago, but people were like, `What's Slack?' I think there was tension created because I'm a little bit more technologically savvy, and I want to use things that I know will help me and I know that have helped others...But their [the blind members] only form of communication is texting...[And,] like the phone call is the ultimate emergency, right? But for a blind person, that's actually the easiest way to communicate.''
\end{quote}
Similar to the trade-offs between Slack and texting the above quote highlighted, participants described pros and cons of a range of tools they have explored from different perspectives. As Board-6 said, ``\textit{everyone brings their own preferences and stuff.}'' 

~\rev{The youth team chose to use Teamsnap, which is a sports management app that has received support most from those who are more experienced in sports club management.} Athlete-3, Athlete-5, and Staff-3 all strongly recommended TeamSnap as the central platform, noting that it is designed specifically for sports management and structured around practice and race schedules, with features such as registration, invoicing, and rostering \cite{teamsnap}. Staff-3, who works as the assistant coach of the youth program, described his preference with the platform based on his prior experience at other clubs:
\begin{quote}
    ``TeamSnap is what I've battled with people for regular, competitive teams. We [his other team] tried a couple different ones, and TeamSnap is relatively inexpensive, or possibly free, and it just worked well.''
\end{quote}
The ``battle'' that Staff-3 described highlights his preferences and priorities from a coach's perspective. ~\rev{As the assistant coach of the youth team, he needs a structured sports management app to help him manage a large volume of information from the team's members and their parents.} Likewise, Athlete-5 advocated for TeamSnap, noting that although the learning curve was steep, it was ultimately worthwhile for streamlining communication for everyone:
\begin{quote}
    ``Once everybody has been there [TeamSnap] for a couple years, everybody understood how it's used. There are so many different aspects of it that cover all those things. You can email from it. You can put documents in folders. You can set up separate chats.''
\end{quote}
However, from Board-2's perspective as a coach of blind members, TeamSnap was still perceived as inaccessible and difficult to use, particularly given that some members are not involved with the organization on a full-time basis. This was supported by Athlete-4, a blind member who used TeamSnap with another club:
\begin{quote}
    ``[TeamSnap] has its own separate pages and all this is bad for the blind. It's not that easy to read all the stuff...You only get the notifications from the page that's open.''
\end{quote}
Given these limitations of TeamSnap to blind members, despite some members expressing strong preferences for TeamSnap, the organization only adopted it for the youth team as all of its members and parents were sighted and coaches primarily communicate with parents to manage RSVPs.

~\rev{Meanwhile, the adult team where the blind athletes belong has been trying to move their communication from texting to WhatsApp, as the coach (Board-2) considered WhatsApp as a middle ground that has a relatively ``straightforward'' interface while allowing channel segmentation and structural messaging.} This point was also supported by Board-6:
\begin{quote}
    ``It just rears its ugly head if everyone goes to texting. When you've got iPhones, and you have iOS and Android devices, curating the list tends to get pretty gnarly. So, the adult team switched to WhatsApp.''
\end{quote}
The numerous devices mentioned in this quote reflect the labor involved in access work. While the operational team members all share a commitment to adapting to members' needs, they must also balance this work with the demands of their own roles and responsibilities. WhatsApp then emerged as a temporary solution, a compromise between accessibility, convenience, and the practical constraints faced by the coaches. Board-6 described WhatsApp's ability to thread messages, which is especially valuable for blind users as it helps screen reader users track conversations:
\begin{quote}
    ``Someone who uses screen reader, they have to read everything. It's always hard to go back to that initial message... If each question has its own post in the text thread, it'd be easier not only for the blind person to be able to find that initial message and reply to it. It's also easier for the person who asked to know that, okay, these are the answers.''
\end{quote}
He also emphasized the usefulness of WhatsApp's channel separation feature to structure information for blind members:
\begin{quote}
``The other thing that we do with WhatsApp is there's a channel there just for one-way, top-down communication. Just announcements, no one asks [questions in this channel]. If you have a question, send a DM... And I'm encouraging people within WhatsApp to create separate chats for specific events. If there's a race coming up, then this particular chat is just for this specific race.''
\end{quote}
Despite these perceived strengths of WhatsApp, the idea of moving the whole organization's communication there still met with hesitation from some members, who felt that the organization was already using too many tools. Additionally, WhatsApp was still a new application for many in the organization and required additional effort for learning and onboarding. For example, Athlete-2 described challenges at every stage, from downloading and setting up WhatsApp to learning how to use it:
\begin{quote}
    ``When [the organization] chose to use WhatsApp, that required everybody on the team to one, download the app and then two, fill out an extensive questionnaire and register, put in the codes for invitation and respond to the individual streams. That can be very challenging for someone like me who is blind.''
\end{quote}
Even though he now found WhatsApp usable most of the time, he still described the application as ``\textit{a little glitchy}''. For example, he mentioned that sometimes new messages were not being read out to him, and he found it difficult to troubleshoot the issue. As a result, he has to develop workarounds, such as \textit{``text somebody, say, hey, what's the new message in the channel.''} ~\rev{The learning effort he described reveals the labor blind members took to gain access, especially those who were unfamiliar with the tool.}

As another middle-ground option, the organization also worked to establish their website as a basic, shared information space, where members could access key details such as practice times, events, and race schedules. Their future plan included posting additional information through blog posts. However, web accessibility of modern, dynamic websites can be as complex as accessibility on the other tools the organization is using. For example, Athlete-4 mentioned that most websites are generally not accessible to blind people, and he described himself ``\textit{not very good at getting on websites at all.}'' Because of this, he ``\textit{never went to [the organization's] website.}'' Maintaining website accessibility can be especially challenging for non-profits due to resource constraints. Staff-1 said that their ``\textit{website is not fully accessible even though it's ADA compliant,}'' while Staff-3 mentioned that ``\textit{the calendar doesn't always consistently get updated,}'' pointing to the calendar's limitations such as technical bugs and the failure to reflect last-minute changes to events.

~\rev{The negotiations over the range of tools and the variance in members' perspectives show that no tool can truly ``solve'' the work of access. Instead, the tools were actively shaping who does that work.} Since there is no single technology that serves as an ideal choice, making communication accessible becomes an ongoing topic of discussion and work for the organization, with staff members putting in intensive labor to repair breakdowns and keep everyone connected. For example, as the website is not accessible to all, the operational team spent considerable time transferring the calendar information on the organization's website to each member’s personal device. In one meeting with the core operational team, we observed how Board-6 and three coaches each spent considerable time transferring the website calendar to their individual devices because of the variety of platforms they used, including Google Calendar, and built-in calendars in iPhone and Android. Similarly, one blind member shared with us on site how Board-6 assisted him in transferring the calendar to his iPhone, showing the one-on-one and time-consuming nature of such efforts. As these day-to-day negotiations and work reveal, members’ disparate needs and preferences not only prompted ongoing reflection on the strengths and limitations of communication tools but also became key sites for collaboration and repair, even though this work requires intensive labor.

\subsubsection{~\rev{Balancing the Distribution of Work}}
While the organization has been experimenting with tools, these practices also prompt reflection on the ~\rev{(re-)distribution of the work of access among all its members. While the organization has prioritized access and inclusion since its founding, the demands of day-to-day work still often push minority disabled members toward less ideal options.} As Board-6 reflected on the tension between accommodating individual needs and maintaining a uniform communication culture,
\begin{quote}
    ``This is always the thing, right? How much accommodation does a person need based on another person who's giving the accommodations assessment of what that person is actually capable of doing and learning? Which sounds kind of bad. But I think, as a disabled person, I kind of have held our blind and low vision people to a little bit of a higher standard.''
\end{quote}
The quote above pointed to a long-standing tension between providing accommodation and assimilation, particularly when it is difficult to meet disabled people's access needs and organizational demands simultaneously~\cite{cohen2010we, apara2025}. During our discussion of this challenge, Board-6 noted that allowing communication decisions to be made solely by end users can be problematic when everyone prioritizes their own preferences:
\begin{quote}
    ``If we left it to voting, and everyone just said text. Then we're still left with the problem of adding new people, taking people out... Going the more democratized route of having people pick, I thought that was just a recipe for disaster.''
\end{quote} 
\rev{The quote above reveals the intensive work involved for the operational team to adapt to individual needs.} Echoing this challenge in balancing people's preferences and work, Athlete-4 mentioned that another club he races with has faced similar difficulties as the organization in establishing a shared communication platform. Like our field site, this club serves cross-disability groups, including members who are BLV, neurodivergent, deaf, and have motor disabilities. To balance differing preferences and opinions, they rely on board members to vote and make the decision:
\begin{quote}
``We have a board of directors who all voted on it. Is Slack the perfect platform? No. Is WhatsApp the perfect platform? No. There's really no perfect platform out there.''
\end{quote}
Athlete-4's experience at a different organization echos Board-6's point that there is no one-size-fits-all solution to collective access in mixed-ability groups. As a result, both organizations implemented mechanisms to guide and shape their communication culture, and balance the work of access for all their members. For our field site, rather than leaving decisions solely to board members, they adopted a middle-ground strategy by delegating decision-making to the coaches of each program. ~\rev{This choice has eased the workload for the coaches of the regular programs, who were considered responsible for most of the communication with program members.} As Board-6 explained,
\begin{quote}
    ``When the coaches realized that they had the power or they had the agency to just set the culture, based on the communication platform, that was the game changer for them.''
\end{quote}
As another way to balance people's preferences, Board-6 also described how he encouraged people to first try and learn to use the tools, rather than defaulting to people's preferences. He commented on the typical challenge for BLV people to find information in information threads this way:
\begin{quote}
    ``Instead of restating what their answer is [in WhatsApp], just holding them [BLV people] accountable to utilizing their own time to find the answer before they ask. It's gotten pushback. But once we were able to correctly determine that the pushback was because of technological skill versus willingness to find the information. Once they learn how to find the information. It settled itself out pretty well.''
\end{quote}
Together, the two quotes above illustrate how the leadership sought to guide the communication culture despite pushback from some members. Yet, as the broader negotiations uncovered in this whole section show, even after adopting a tool or practice, the team must continually adapt to meet members’ evolving needs and the technical limits of each tool. These ongoing efforts highlight how access is a continuous process, in which the team identifies who might be marginalized and works to keep everyone connected.

\subsection{Navigating Heterogeneous Roles, Expectations, and Communication Norms}
\rev{Our conversations with the organization members about their use of communication tools surfaced additional axes of power asymmetry beyond disability.} Participants' reflections on the challenges in organizing communication made it clear that the use of tools was complicated not only by varying access needs, but also by members' varied roles and the communication expectations and social norms tied to them. When we asked Staff-3 for suggestions on communication tools, he emphasized that clearer norms and expectations were the foremost needs:
\begin{quote}
    ``Every tool has its pros and cons, but it's a bigger issue if the communication isn't happening in the first place... You can't fix that by `let's use WhatsApp now.' It doesn't matter what the communication method is... Sometimes it was on this app, sometimes it was in this text, sometimes that email, sometimes it was this phone call, but it's all communicating.'' 
\end{quote}
The point that ``You can't fix that by `let's use WhatsApp now,''' suggests that for him, the greater need is not simply finding and using a tool, but creating a clear communication flow that accommodates members’ diverse roles, types of involvement, and communication expectations. For any communication technology to function effectively, it must be aligned with people's actual work patterns and practices~\cite{garg2023orchestration, tang2015restructuring, conway1968committees}. However, finding this alignment itself is difficult in a multi-purpose organization that combines professional and leisure-oriented participation and encompasses varied roles (e.g., paid staff, volunteers, parents), often leaving the work of navigating these differences on individuals in the leadership or operation team, or those who are marginalized by existing communication norms.

\subsubsection{Varying Degrees of Engagement} 
A key source of complication lies in varying degrees of engagement among the members, which can cause persistent gaps in communication, particularly between those who are more closely connected to the organization and those who are not. Many of the adult members, for example, began as a small group competing together and grew into a close-knit network that has known each other and been involved with the organization for years, some even for decades. These long-standing interpersonal relationships, the day-to-day experience of physically being together, and a strong, casual vibe rooted in the sports culture provide rich opportunities to keep people stay connected and informed, especially for those who are less tech-savvy or have limited access to digital communication tools. However, reliance on in-person communication can marginalize those who are less connected, such as parents who are newer to the organization and unable to attend frequently due to work obligations. These varied relationships also result in uneven distribution of labor, placing additional responsibilities on certain individuals, such as the staff members who are more involved in day-to-day operations and are expected to relay information across members.

For example, the most active athlete members tend to have greater communication through both in-person and direct messaging about activities and events, as they have developed close relationships with the core members of the operating team. Athlete-4, who is familiar with the leadership team as racing with them and trying out their new devices, described the acceptance, if not direct encouragement, of informal communication within the organization, \emph{``They [the operating team] have always said, `If you have a question, ask us'. They never had a problem with me calling up and asking a question.''} Similarly, several other highly engaged members found that the most convenient and comfortable way to reach out to the coaches was through direct messages on their personal social media accounts, rather than through TeamSnap or other platforms designated by the organization for broader communication. While these practices help individual community members bridge gaps in communication, they add to the work of organizational staff, who must answer individual questions that are not only coming to them through channels the organization uses (e.g., text, phone calls, Slack, WhatsApp, TeamSnap) but now also through their personal social media accounts. 

Communication about events and volunteer recruitment also often occurs informally, yet these messages might be missed in remote channels. On-site, several members shared that they often hear about events informally, through simple exchanges like, ``\textit{Hey, there's an open event next week that really needs help. Are you comfortable with that?}'' suggesting that event announcements and arranging volunteer support are often done through in-person word-of-mouth. In contrast, members who are less connected to the group may miss information scattered across the channels that the organization uses. As Board-6 described, a key challenge the operational team faces involves disseminating information:
\begin{quote}
    ``If we have a big event on Saturday, we'll post something in Slack, and then we ask each program leader to disseminate that information to their group. But it's hard to figure out if people are getting it.''
\end{quote}
This way, the variations in involvement can make reliance on interpersonal communication (both in-person and through direct messaging) a double-edged sword. While active members stay informed through ongoing interactions and benefit from the close relationships, those who are less connected to the community may feel lost or unsure about where to find the information they need. For instance, Board-6 described a key difference between the athletes and parents in terms of engagement:
\begin{quote}
    ``The program that uses TeamSnap [the youth program] is definitely more like the Little League kind of parent, and everyone's going crazy because they're in 5 or 6 different things, whereas the adult programs, you know, their focus really is on [the sport], and so they're a little bit more engaged on that.''
\end{quote}
As illustrated by this quote, parents of children involved in the organization often juggle multiple schedules and communication channels for their children, alongside the demands of their jobs. For example, we observed that those who tend to stay on site during sessions often did not work outside the home, allowing for more flexibility in their daily routines and opportunities for informal communication with others at the organization. In contrast, one parent we interviewed is unable to come to the site and instead has an assistant drive their child to the organization. This parent expressed confusion about how the organization is structured and the roles and responsibilities of its members, particularly those who work remotely or are rarely on-site, such as whether the financial accountant could assist with processing refunds. This parent was also surprised to find that they missed all the public events, even though the team had published all the events on their website calendar and social media. Although the parent knew they could reach out to the team with questions, they expressed hesitation about repeatedly contacting the leadership. As they commented:
\begin{quote}
    ``There's no person to contact other than the leader. So, I always feel bad about bugging him. He's running all this stuff. So I really hesitate so much. I just kind of don't contact him.''
\end{quote}
The feeling this parent described highlights a common disconnect in communication expectations that stem from differing levels of engagement among members. Even when the staff is open and willing to respond to clients' questions, some may hesitate to reach out due to social norms or perceived boundaries. As a result of these misalignments in communication expectations and access to information, staying involved becomes ongoing relational work, and it might often feel like an individual rather than collective responsibility.

\subsubsection{Mixed Working Models and Expectations}
The organization also contends with varying levels of involvement among those who support day-to-day operations, including organizational leaders, paid staff, coaches, therapists, and volunteers who collectively manage programs, activities, and events. These diverse and often overlapping roles result in mixed work models and expectations for communication. This challenge is evident in the gap between leadership’s idealized vision of streamlining the workflow for the whole organization and the practical complexities of navigating this work. As Board-6 described, ideally, the organization can use Slack and Asana to divide tasks and communication channels:
\begin{quote}
    ``We're using Slack as our internal leadership communication platform. And then we've hooked into Asana so that we can just set tasks for people... I think it's just easier, and everyone agrees that just make them an Asana task and just assign it, they get the notification, no need to do that, extra communication step. And then, one step down from that, or when you're outside of the management and leadership curtain, each program is allowed to utilize whatever communication platform that they want.'''
\end{quote}
In this model, people involved in leadership and operations are also expected to bring information from other channels (e.g., text messages with individual members, group chats in WhatsApp) into the Slack workspace as a way of sharing information with other members at this level. As Board-2 described:
\begin{quote}
    ``I think a lot of conversations start in texting, because it's, I just have my phone and I'll just text someone, and then one of us will say, `Hey, you know what? That's good information that should be shared to everybody. We should probably move that to Slack,' and we'll copy and paste it into a Slack channel, so that other people can see what has been talked about.''
\end{quote}
The above two quotes illustrate the typical flow of information practiced within the organization and how communication is expected to be divided between smaller groups communicating through various tools and channels that work well for them (e.g., WhatsApp, TeamSnap, text messages) and a shared Slack workspace for people who support operations and leadership. However, in this assumed process, the labor of transferring information across platforms might often fall on operations and leadership staff, as they are expected to manually relay information across different groups and levels within the organization.

Moreover, while the intention behind this fragmentation is to enable smaller groups to communicate in ways that work well for their idiosyncratic needs, challenges arise when effective communication across the organization depends on individual people maintaining awareness of information and moving it across varying groups and platforms. An underlying issue is that people within the organization have varying orientations towards their work, with some being paid staff and others being volunteers, who need flexibility and may not be able to engage as regularly as others members of the organization. One participant, for example, described how they sometimes felt lost in coordination efforts, reflecting the confusion that arises when communication paths are unclear. As they said:
\begin{quote}
    ``Many times I was asked to follow up with a list of people, but I didn't get all the phone numbers for them... I wasn't even sure how much of that I was supposed to do...[It turns out] I was doing X but that wasn't necessarily what they told me I was going to be focusing on. My role was, like I said, very fluid. I don't mind it being fluid, but it's kind of tough when it's like, `oh, I guess this is what I'm doing now.'''
\end{quote}
Similar to other non-profits, many people at the organization occupy multiple roles and volunteer wherever they are needed~\cite{netting2005mixing, lopez2020inside}. For example, while Staff-3 officially serves as the assistant coach of the youth team, he also helps with other events and responds to needs like piloting adaptive equipment. While many people understand that such volunteer work is fluid and flexible, communication breakdowns impact their expectations regarding volunteering and some members' access needs. For example, one participant said,
\begin{quote}
``It's hard to schedule my calendar around things that are not clear. I think they might miss out on additional people who want to help and how they can help. I think a lot of people would help more if it was clearer.''
\end{quote}
Similarly, participants who have clear access needs might desire clarity when it involves scheduling their bus services or preparing assistive equipment accordingly. As one said,
\begin{quote}
``Lots of the members are blind and they have to arrange a ride to get somewhere. Do you want me to serve food and be on land, or do you want me to be on a boat? I need to plan and know what I'm gonna do. I just didn't feel like I was getting the information I needed in time to make those decisions. And then I felt guilty, or felt like I had done something wrong because I wasn't showing up or supporting the team.''
\end{quote}
Together, these two quotes reveal how the responsibility for navigating ambiguity in volunteer roles could fall on individuals due to the organization’s fragmented communication setup. This disconnect, in turn, can reduce motivation to take on volunteer work and make structuring and coordinating communication even more difficult.

\subsection{Adapting to Shifting Organizational Demands from Cross-Disability Expansion}
The organization's evolution from a small group to a larger professional organization, combined with cross-disability dynamics that shift with the distribution of disabilities, adds more complexities in accessibility considerations. A major challenge stems from the need to respond to rigid policies related to serving disabled populations. Board-1 described that the organization has been struggling to streamline information across communication channels while remaining compliant with the U.S. Health Insurance Portability and Accountability Act (HIPAA), as disability status and associated care details are are private information.
\begin{quote}
    ``It got hard adding people to the [Slack] channels, and then keeping up with who was actually staying on as a volunteer and who is not...When we're looking at participant information, because of HIPAA guidelines and protecting client identity, we have to be careful about what we are sharing [and] who we're sharing it with.''
\end{quote}
The quote above shows how the lack of HIPAA-compliant tools further restricted the organization's organizing of communication and personnel. Although the leadership attempted to maintain a regular volunteer team , keeping a team of volunteers with the necessary expertise remains challenging.

Expanding into a cross-disability organization also prompted new conversations around whose needs are prioritized. These complexities are reflected in the organization's distribution of resources and focus of attention (such as physical equipment and outreach strategies), which collectively shape who feels welcome to join and remain engaged. The staff team frequently mentioned the challenges in adapting their equipment and services to cross-disability groups. For example, Board-4 shared they failed to reach deaf communities after a negative encounter with a deaf person who was interested in coming to their open events. As he recalled:
\begin{quote}
    ``We failed miserably with the deaf community. There was an opportunity. Somebody had heard about us who was deaf, reached out. We had never interacted with somebody from that community before. I remember we asked if they could provide an interpreter, and the response was, `How dare you ask that? You have to provide an interpreter.'''
\end{quote}
In this instance, the tension arises from conflicting expectations about who is responsible for providing accommodations, the team's limited experience working with deaf community members, and the constraints on resources for hiring interpreters. Similarly, Board-3 and Board-5 noted that, due to funding limitations, they must prioritize investments in physical equipment, leaving accessibility largely reactive to the needs of existing members. While choices of which equipment to purchase might not be directly related to `communication technology,' such choices can influence who feels like they belong or not.

The complexities in reaching cross-disability groups are also reflected in the challenges in outreach. As there is no single platform or method that is accessible across disabilities, anticipating attendance and organizing accordingly becomes hard when planning public events. As Staff-1 said,
\begin{quote}
    ``We just don't have a solid strategy on outreach... I think the more we look at incorporating things like the blog posting on our website or trying new email campaigns, there's just always a hesitation, because we don't know if those things are accessible.''
\end{quote}
As described in this quote, despite the organization's efforts to disseminate information across platforms including YouTube, Instagram, X, BlueSky, and others, there remains a persistent uncertainty about the effectiveness of these strategies. While the organization has taken steps to improve accessibility, such as ``developing audio descriptions for YouTube videos, and adding subtitles'', much of its outreach and promotion continues to rely heavily on word-of-mouth or offline campaigns, making it difficult to anticipate headcounts for their public events unless the events are with specific organizations. These uncertainties add yet another layer of complexity to predicting who is likely to participate and coordinating accordingly.

The disability distribution is further complicated by structural factors associated with different disability groups. For example, even though the organization was aimed at attracting anyone from any disabled community who was interested in getting involved, the organization has begun serving an increasing number of youth with intellectual and developmental disabilities, leading some adult members to wonder whether that has become the primary focus of the organization. When we asked about Board-1 how the organization is having an increasing number of children who are autistic or have developmental disabilities, she explained that the outcome might be affected by the distribution of disability in general:
\begin{quote}
    ``I would just say the prevalence of those types of disabilities are highest right now in our country... So I think that's why we get a lot of children that have that type of diagnosis.''
\end{quote}
Her explanation reflects how broader dynamics of cross-disability work to shape the participation at the organization. Even when an organization strives to be inclusive of all disabilities, such efforts can unintentionally create imbalances in attention or resource allocation. As yet another factor influencing the cross-disability dynamics, board members often mentioned the influence of revenue for providing services to certain groups (i.e., autistic youth) and funding over what programs they can offer (Board-3, Board-5, Board-6). While the organization actively seeks funding for cross-disability activities, much of it depends on who participates. As Board-6 explained,
\begin{quote}
    ``As we move forward, a lot of people in our regular program, they come to us through [a local disability center], so they're bringing funding from that center, which, in turn, has allowed us to at least have some stable income.''
\end{quote}
Echoing this quote, we heard from many parents that they received funding from the center Board-6 mentioned to participate in the program. For example, Parent-2 shared that the center quickly approved their request to join, recognizing the program's potential benefits for autistic children, particularly in offering opportunities to interact with a diverse range of people. As such, the availability of funding and revenue for each sub-community may introduce influences over which disabilities are more represented within the organization.

Given the pressure to react to shifting participation and scale accessibility for cross-disability groups, Board-3 shared that the organization is intended to keep its size small, aiming instead to serve as a model for other clubs. This strategy not only reduces the pressure to scale accessibility internally but also acknowledges the reality that no single organization can address every challenge alone. Board-3 shared with us the story of one member they served, highlighting the value of expanding their model, not within the organization itself but externally:
\begin{quote}
``We have a really high-performing athlete who comes and drives from [a far place]. She's amazing, but you wonder if there's a club that would be closer geographically to her. I think, what we would love to see, is that there be more of these type of programs.''
\end{quote}
The cross-organizational perspective reflected in this quote highlights that access should be treated as an ongoing work across the society. While making a single organization fully accessible to everyone is inherently challenging, the ideal scenario envisioned by organizational leaders emphasizes that disabled individuals should have access to a variety of spaces that support their diverse needs and goals. It is in this vision of collective access that the cross-disability and mixed-ability model truly demonstrates its potential and value.

%% file: sections/05discussion.tex
Our analysis foregrounds the ongoing work of how access is ~\rev{practiced} within a mixed-ability organization and the complexity of designing `solutions' to conflicting access needs. Drawing on our findings, we argue for the need to shift from viewing ``conflicting access needs'' as problems to be solved through design to generative sites for reflection and collaboration in access-making. Below we draw on work from disability studies and disability activists to discuss how this conceptual shift is essential for designing for collective access. We then provide recommendations to support this shift in HCI research and design.

\subsection{From Designing for ``Everyone'' to Designing for Collective Access}
The notion of conflicting access needs foregrounds the criticality of designing for a diversity of (dis)abled users. Hence, a natural question is to ask ``Why can't we just design for everyone from the start?'' Indeed, Universal Design (UD), an approach to creating products and environments that are usable by as many people as possible without adaptation, is often viewed as a practical response to the need to design for diverse bodies and minds. Yet, disability studies scholars have long critiqued the false promises and problematic logic of UD~\cite{hamraie2017building, hamraie2012universal, titchkosky2011question, dolmage-ableism}. Aimi Hamraie, for instance, argues that the emphasis on universality can exclude people with more complex and marginalized needs and hide the politics embedded in design choices~\cite{hamraie2017building}. Their in-depth analysis of UD shows how ``design for all'' still often is designing for particular bodies while excluding others, and how designers conceptualize ``everyone'' depended on whether our presence had been anticipated, in what ways, and for what purposes... When the goal is to design for `everyone,' I ask, who counts as everyone and how do designers know?'' Jay Dolmage similarly argues for an anticipatory stance, saying ``The push toward the universal is a push toward seeing space as open to multiple possibilities, as being in process,''~\citep[p117]{dolmage-ableism}. He further cautions that ``turning UD into a checklist defeats so much of the rhetorical purpose of UD, as what I have called a `way to move,'''~\citep[p145]{dolmage-ableism}. Echoing their points, our analysis shows that designing for ``everyone'' within a mixed-ability organization is not simply to identify the shared needs or seek universal solutions. Rather, it involves continual attention and adjustment to the dynamic configuration of specific technological features, people, and structures as it is about the ongoing reflection and collaboration that emerges through the working out of collective access. Designing for collective access, therefore, is an ongoing situated work that is performed locally by those most affected.

Building upon this ongoing, situated notion of access, our findings further show that what are often framed as conflicting access needs are not simply a `problem' to address, but rather critical sites for examining the politics of access and pushing the ongoing work of access forward. As our findings show, the conflicts encountered at our field site are crucial for revealing dynamic, situational dependencies embedded within broader organizing structures. The conversations around members' diverse needs have led to reflections over technical constraints, organizational ~\rev{hierarchies, loopholes, and competing demands, while directing attention to members who are marginalized or take on extra labor to maintain access, such as BLV members who lack access to the shared workspace, new members who lack access to community information, or coaches who invest significant effort in transferring information and connecting people.} In this way, tensions arising from ``conflicting access needs'' reflect pervasive power asymmetries embedded within a broader network of tools, interpersonal dynamics, and socio-political forces operating at multiple scales. Our findings therefore draw attention to the need for constant reflection on the structural and political conditions that produce these ``conflicts,'' instead of locating their resolution in individual responsibility or flawed technology design. In framing conflicts as a generative space, our findings also question the assertion that ``conflicting access needs'' \emph{should} be resolved through developing shared norms, even if that is what happens in practice~\cite{mack2021mixed, yildiz2023virtual}. As seen in the negotiations over tool choices at our field sites, as most members organization emphasizing collective norms can pressure people with less power such as the BLV athletes to conform by learning and adopting new tools, even though the fundamental sources of this pressure may be an industry-wide focus on visual-centric products to support organizational work or ingrained ableist expectations of perseverance through inaccessibility. To foster ongoing reflection on these power asymmetries and broader politics of access, we argue that conflicts should be seen as opportunities to repair relations within inherently uneven systems. A closely related idea is what disability activist Leah Lakshmi Piepzna-Samarasinha calls the principle of mutual aid~\cite{leah-collective}, which emphasizes accessibility has no perfect solution and emphasizing the ongoing, sometimes messy practice of supporting one another. Drawing on both our findings and these ongoing reflections in critical disability scholarship and among activists, we view ``conflicts'' as constitutive of the continuous work of access, inviting reflection and collective action. Below, we discuss implications for future HCI research and design of collective access.

\subsection{Implications for Research and Design}
We offer two ~\rev{implications} for future research and design on collective access.

\subsubsection{Treating conflicts as sites to reveal existing power structures}
~\rev{A key way to shift ``conflicts'' from problems to solve to generative sites is to treat them as opportunities to reveal existing power structures that marginalize disability.} Many of what are called ``conflicts'' in access emerge not from incompatible needs, but from the ways systems are organized to marginalize disabled bodyminds. For example, in our case study, tensions between BLV members' needs for non-visual communication and the organization's desire for more structured collaborative technologies reflect systemic neglect of non-visual interaction in design rather than irreconcilable access needs. ~\rev{Similarly, the root causes of many conflicts in meeting access needs arise from systemic marginalizing forces, not only inaccessible tools, but also rigid policies~\cite{moore2025executive} and broader cultural and economic systems, such as elitism~\cite{zhang2026access}, neoliberalism~\cite{ly2025accessibility}, and capitalism~\cite{apara2025}. When conflicts are treated as problems to be solved, the push for a ``solution'' often risks pressuring individuals with less power, often disabled people, to conform, leading to hidden and unrecognized labor such as learning to use new tools~\cite{branham2015invisible, yildiz2023virtual, price2024crip, zhang2026access}. To shift these dynamics, conflicts should be treated as critical opportunities to surface hidden forms of labor and underlying power structures, making them visible rather than simply ``solved,'' and opening them up for collective reflection.}

A crucial next step for research is to gain a deeper understanding of how collective access is enacted and the political dynamics at play across diverse contexts. As a recent systematic review of accessibility research by Mack et al. shows, much of the existing work focused on a single disability~\cite{mack-sys-review}. From 2010 and 2019, more than 70\% of the accessibility studies published at CHI and ASSETS focused exclusively on single disabled groups~\cite{mack-sys-review}, leaving limited understanding of how accessibility should be practiced and sustained in complex mixed-ability situations. While an increasing body of work has incorporated lens of power in understanding the adoption and use of assistive technologies~\cite{hsueh2025minor, das2019doesn}, the shaping of access~\cite{das2019doesn, mack2021mixed}, or power dynamics in large institutions~\cite{ly2025accessibility, apara2025}, much work remains to do to uncover collective access practices and politics in diverse organizational contexts. For example, much of the existing work has relied on individual perspectives such as disabled workers~\cite{das2019doesn, apara2025} or accessibility coordinators~\cite{ly2025accessibility}, making a multi-stakeholder understanding a key gap in the literature.  While the present study of a non-profit organization begins to address this gap, more situated studies of the working out of collective access are essential as access practices are uniquely shaped, and often constrained, by specific institutional arrangements. ~\rev{For example, the access in our field site is uniquely shaped by its memberships and organizational hierarchies. Most staff members identify as non-disabled, and most athletes are adults who are BLV or have physical disabilities or are children who have neurodevelopmental disabilities. The discussion over communication accessibility was therefore mostly around accommodating the blind members' preferences for non-visual communication methods and influenced by the power non-disabled staff members and coaches have.}

We also urge future research to diversify methods, not to resolve, but to spark dialogues around the politics of collective access, such as exploring scenarios that question who is included or excluded in design decisions, or imagining systems that actively surface and shift power dynamics. The systematic review Mack et al. conducted shows that 94.3\% of the papers they analyzed focused on user studies, with 84.1\% of them drawing on interviews, usability testing, or controlled experiments as their methods~\cite{mack-sys-review}. In contrast to these studies that prioritize immediate user feedback or performance metrics, we advocate speculative design as a promising yet relatively under-explored set of methods within accessible computing to provoke design around power. In speculative design, designers conceive imagined scenarios to question the assumptions, values, and power structures built into existing systems~\cite{auger2013speculative}. It is therefore distinct and even opposed to direct problem solving~\cite{auger2013speculative}. For example, prior work has employed speculative methods to understand neurodivergent people's ideal of emotion technologies~\cite{zolyomi2024emotion}, envision deaf-first technologies~\cite{angelini2024deaf}, and to provoke reflection on normative assumptions in the design of assistive technologies~\cite{spiel2022transreal}. ~\rev{Similarly, speculative methods can be used to prompt reflection on access in diverse mixed-ability contexts. For example, for our study, imagining variations in organization size and the distribution of disabilities may help people surface the trade-offs in communication technology adoption and reflect more deeply on the politics involved.}

\subsubsection{Treating conflicts as sites of accountability and repair}
~\rev{Another way to move beyond a problem-solving mindset is to approach conflicts as productive sites for accountability and repair. Equally important to exploring norms of collective access~\cite{yildiz2023virtual, mack2021mixed, alharbi2023accessibility} is ensuring they are accountable, for example, by making decision-making processes transparent. As a step further, conflicts should serve as key moments to continuously reflect on harms inflicted on members and to explore ways to repair them. In the case of our study, to prevent BLV members from being marginalized after the adoption of Slack, the organization's leadership and operations continue to engage members through day-to-day interactions and active information sharing, such as transferring the website calendar to each member’s device, and creating custom communication spaces across texting groups, Slack channels, and WhatsApp groups. Similarly, research found that workers in mixed-ability teams often undertook intensive work to transform information and documents into accessible formats, while this work is often invisible~\cite{zhao2026accessibility}. These hidden forms of labor, which generate new forms of uneven distributions of work, should become a new site for accountability and repair. For example, organizations should develop ways to recognize these organizational work important to accessibility. Meanwhile, design should focus on minimizing access labor, such as developing tools to help people curate information across platforms and support cross-platform message forwarding.}

~\rev{While our study focused on mixed-ability work in a non-profit, the insights from this research have broader relevance to designing for collective access in other contexts. For example, ``conflicting values'' emerge as a key challenge to developing visual assistive AI systems for BLV people~\cite{hanley2021computer}, because BLV people's visual access needs often ``conflict'' with other values such as bystanders' privacy and risk of making inferences from visual content (e.g., race and gender)~\cite{tang2025everyday, bennett2021complicated}. While AI research often emphasizes improving model accuracy or de-biasing methods, treating ``conflicts'' as key sites for accountability and repair shifts focus to designing systems that provide reliable and responsible access. As Meredith Moore, a scholar who has ADHD,  commented on technologies she wants,}
\begin{quote}
    ``I didn’t need perfect personalization or disability-specific tools. I needed systems that made space for imperfection, invited re-entry after failure, and assumed users are doing their best with the capacity they have. The key design question isn't whether a system works -- but who it works for.''~\cite{moore2025executive}
\end{quote}
The emphasis on who technology works for in the above quote echoes the principle of mutual aid~\cite{leah-collective} and the care work of access, i.e., the continual work of supporting each other~\cite{bennett2020care}. To support such ongoing work of access making, conflicting values and needs AI systems face should shift from problems to be solved to key sites for designing mechanisms for ongoing collaboration and repair. For example, AI systems should incorporate back-up or alternative access options, such as human assistance or community support, to ensure that disabled people can receive continuous care. They should also develop repair strategies after harm occurs, such as acknowledging harm, attributing responsibility, providing remedies, and enabling systemic change~\cite{xiao2025comes}. Treating conflicts as sites of accountability and repair opens up many of such rich design possibilities to turn access into care work.

\subsection{Limitations}
This study has several limitations. First, while we incorporate both interview and observational data to enhance our understanding of the organization, our understanding is limited to the time and spaces in which we were able to be present. For example, we were unable to directly observe internal communication channels or personal text exchanges, which were described as a key part of communication. Second, our findings are limited to the perspectives of participants who were willing to participate in our interviews ~\rev{or we engaged with on the site}. The findings may not fully represent the views of all members within the organization, as those who were willing to participate may have taken a greater interest in issues central to our study or came to the site frequently. ~\rev{Additionally, out of ethical considerations, we did not include perspectives from adults with neurodevelopmental disabilities who were unable to provide informed consent and children under 18. Our findings might therefore miss their perspectives on what is collective access.} Third, any such account researchers generate of others is partial and must be read alongside other studies that attempt to capture the lived realities of its informants, further underscoring the need for more empirical and theoretical work on collective access.